\documentclass[preprint, review, 12pt]{elsarticle}
\usepackage{graphicx}
\usepackage{amsmath,amssymb,amsfonts}

\journal{Physics Letters A}
\begin{document}
\begin{frontmatter}
\title{Quantum Illumination with a Parametrically Amplified Idler}
\author{Jonathan N. Blakely}
\address{U. S. Army Combat Capabilities Development Command Aviation \& Missile Center, Redstone Arsenal, Alabama 35898, USA}

\begin{keyword}
quantum illumination \sep second order interference \sep quantum radar \sep quantum  hypothesis testing \sep phase-insensitive cross correlation
\end{keyword}

\begin{abstract}
Quantum illumination uses a quantum state of the electromagnetic field to detect the presence of a target against a bright background more sensitively than any classical state. Most often, the quantum state is a two-mode squeezed vacuum consisting of signal and idler modes with a non-zero phase-sensitive cross correlation, which serves as the signature for target detection, and a zero phase-insensitive cross correlation, which means the modes produce no fringes in second order interference. Here it is shown that applying phase-sensitive amplification to the idler modes of a two-mode squeezed vacuum results in a non-zero  phase-insensitive cross correlation enabling reception by a simple beam splitter and photodetectors. It is shown that quantum illumination with a parametrically amplified idler has a lower probability of error than an asymptotically optimal classical-state scheme in discriminating between a present target and an absent target with equal prior probabilities. 

\end{abstract}
\end{frontmatter}

\section{Introduction}
\label{Intro}
Using a quantum state of the electromagnetic field as a probe, the presence of a target can be detected with lower probability of error than with any classical state of equal energy \cite{lloyd2008enhanced, tan2008quantum}. This approach, known as quantum illumination, exploits entanglement between photons transmitted and photons retained, even though this entanglement is destroyed by lossy reflection from the target and the influence of background noise.  Quantum illumination has been the subject of extensive theoretical study \cite{shapiro2009quantum, guha2009gaussian, ragy2014quantifying, zhang2014quantum_b, weedbrook2016discord, wilde2017gaussian,zhuang2017optimum, zhuang2017entanglement, xiong2017improve, bradshaw2017overarching,  de2018minimum,  nair2020fundamental, Karsa2020Noisy, Karsa2020Quantum} and has been demonstrated experimentally \cite{zhang2015entanglement}. However, quantum illumination has not yet formed the basis of a practical quantum sensing technology as several barriers exist to convenient implementations \cite{shapiro2020quantum}. Nonetheless, quantum illumination is fundamentally important as the first example of a bosonic system that provides a quantum advantage in an entanglement-breaking channel subject to an energy constraint. Thus it is important to consider variations on the scheme to better understand quantum sensing in noisy, lossy environments, and to overcome the barriers to practical application \cite{zhang2014quantum, barzanjeh2014quantum, sheng2015quantum, zhuang2017quantum, liu2017discrete, lanzagorta2018virtual, fan2018quantum, jo2020quantum}.

In that spirit, this article examines the consequences of applying parametric amplification to the retained field modes (i.e. the idler). The parametric amplification considered here is phase sensitive in that it increases one quadrature component of the field while decreasing the other. This operation is shown to change the nature of the correlations between the signal and idler modes such that they become measureable in second order interference. Thus the quantum illumination receiver can be a simple combination of a beam splitter and two photodetectors. 

\begin{figure}[tbh]
\includegraphics{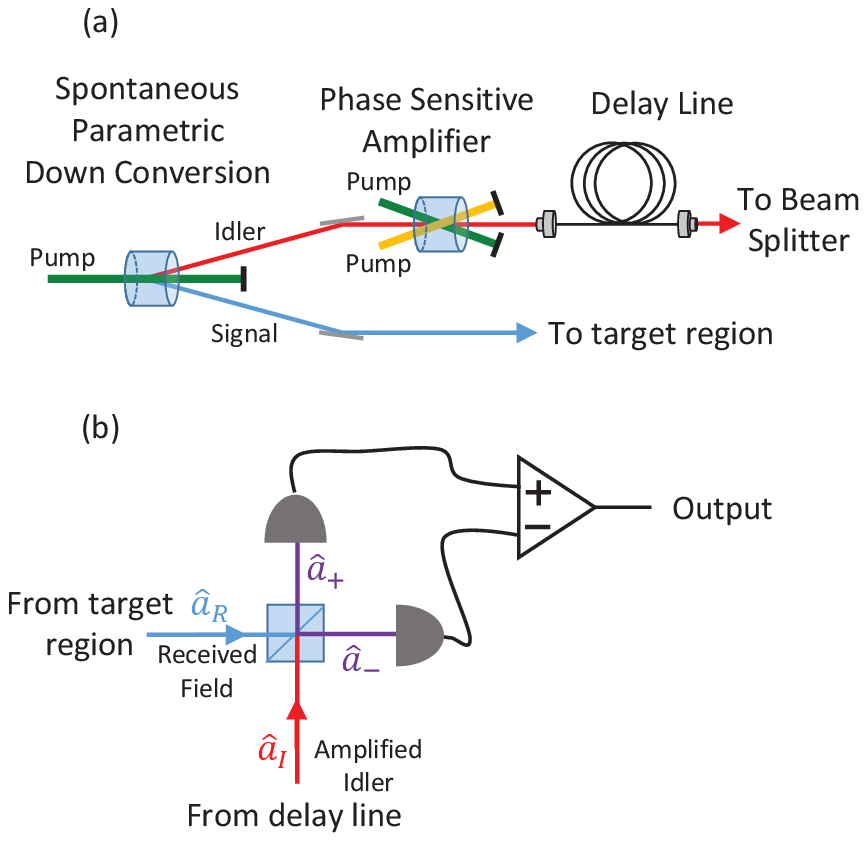}
\caption{(Color online) Schematic diagram of quantum illumination with a parametrically amplified idler. (a) The source is a spontaneous parametric down converter that produces signal and idler modes.The signal mode directly interrogates a region of interest. The idler is input to a phase-sensitive amplifier and delayed. (b) The amplified and delayed idler is combined on a beam splitter with radiation received from the region of interest. Photodetectors at each output port count photons and the difference signal is formed for comparison with the decision threshold.}
\label{fig:Schematic Diagram}
\end{figure}
This variation on the quantum illumination theme is depicted schematically in Fig. \ref{fig:Schematic Diagram}. A spontaneous parametric down conversion source, here depicted as a strong pump laser driving a nonlinear medium (see Fig. \ref{fig:Schematic Diagram}(a)), generates entangled signal and idler beams. The signal beam propagates to the target region. The idler passes through a second pumped nonlinear medium where it undergoes parametric amplification before entering a delay line for storage. Due to the inclusion of the phase-sensitive amplfication stage, the receiver architecture (see Fig. \ref{fig:Schematic Diagram}(b)) simply consists of a beam splitter and photodetectors configured much like a Brown-Twiss interferometer \cite{loudon2000quantum}. The optical implementation shown here is inspired by a recent experiment demonstrating phase-sensitive amplification of one of two correlated beams produced by four-wave mixing in a rubidium vapor cell \cite{li2017improved}. Implementation at microwave frequencies is also conceivable. Superconducting parametric converters and amplifiers for microwave signal generation are now widely used \cite{chang2019quantum, barzanjeh2020microwave}, and microwave Hanbury Brown-Twiss experiments have been conducted \cite{chen2011microwave, peng2016tuneable}.

In Sec. \ref{TMSVcorrelations}, the quantum state of the entangled fields generated by spontaneous parametric down conversion and the correlations between them, which serve as the signature of target presence, are described. In Sec.\ref{AmplifiedIdler}, the effects of parametric amplification of the idler beam on these correlations and their detection are explained. The non-separability of the state is verified. In Sec. \ref{QuantumIllumination}, this state is used as the input to the binary-hypothesis test commonly used to model quantum illumination. In Sec. \ref{QuantumAdvantage}, the probability of error in target detection with equal prior probabilities is derived and compared to a classical benchmark to establish the regime of quantum advantage. In Sec. \ref{Discussion}, the significance of these results is discussed.

\section{Correlations in the Two-Mode Squeezed Vacuum State}
\label{TMSVcorrelations}
The transmitted and retained fields in quantum illumination are entangled pulses generated by the process of spontaneous parametric down conversion \cite{shapiro2020quantum}. In this nonlinear process, photons from a strong pump pulse are converted into pairs of photons, one each in a signal mode and an idler mode, eventually producing spatially separated, correlated signal and idler pulses. Spontaneous parametric down conversion can occur at visible wavelengths, for example in certain crystals or atomic vapors, or at microwave wavelengths, in cryogenically cooled, superconducting circuits. Conversion is only efficient over a limited bandwidth. Typically, it is assumed that down conversion efficiency is roughly flat across a frequency band of width $W$, and that all other processes can be neglected within this band. Each pulse is assumed to be of duration $T$ and contain $M=TW$ temporal modes  \cite{shapiro2020quantum}. Thus, $M$ is both the time-bandwidth product  and the number of modes in the signal and idler pulses. Alternatively, the bandwidth can be narrowed to $1/T$ to allow only a single temporal mode. In this case, the time-bandwidth product is made equal to the same $M$ by transmitting $M$ single-temporal-mode pulses, each of duration $T$, in a total time of $MT$. This approach will be necessary in order to implement the amplification and reception scheme described below.

Each single mode in either approach is characterized by the same mean photon number $N_S$. Also, each signal mode is entangled with exactly one idler mode, and independent of all others, in the two-mode quantum state known as a two-mode squeezed vacuum. This Gaussian state can be represented by the Wigner distribution covariance matrix \cite{weedbrook2012gaussian}
\begin{align}
\mbox{\boldmath$V$} = \frac{1}{2}
\begin{pmatrix} 
\nu  & 0 & c  & 0 \\
0 & \nu  & 0 & -c  \\
c   & 0  & \nu &0 \\
0 & -c  & 0& \nu  
\end{pmatrix}
\label{TMSVcov}
\end{align}
where $\nu = 2N_S+1$ and $c = 2\sqrt{N_S(N_S+1)}$. 

The off-diagonal elements of a  two-mode covariance matrix determine the correlations between the modes. Two types of correlations occur which may be distinguished by their response to a phase shift $\theta$ applied to both modes \cite{shapiro2020quantum}. The \em phase-insensitive cross correlation \em between two modes $\hat{a}_1$ and $\hat{a}_2$, given by $ \left \langle \hat{a}_1^\dagger \hat{a}_2 \right \rangle$, is invariant under a phase shift.  In terms of the elements of the covariance matrix, this correlation can be expressed as
 \begin{align}
\left \langle \hat{a}_1^\dagger \hat{a}_2 \right \rangle =\frac{1}{2}\left \{ (\mbox{\boldmath$V$})_{13} + (\mbox{\boldmath$V$})_{24} + i\left [(\mbox{\boldmath$V$})_{14}-(\mbox{\boldmath$V$})_{23} \right ] \right \}
\label{PICC}
\end{align} 
The \em phase-sensitive cross correlation \em  is $\left \langle \hat{a}_1 \hat{a}_2 \right \rangle$. This correlation changes in phase by $2 \theta$ when the modes are shifted by $\theta$. In terms of the elements of the covariance matrix.
\begin{align}
\left \langle \hat{a}_1 \hat{a}_2 \right \rangle =\frac{1}{2}\left \{ (\mbox{\boldmath$V$})_{13} - (\mbox{\boldmath$V$})_{24} + i\left [(\mbox{\boldmath$V$})_{14} +(\mbox{\boldmath$V$})_{23}  \right ]\right \}
\label{PSCC}
\end{align}
From Eq. (\ref{TMSVcov}), the two-mode squeezed vacuum has phase-sensitive cross correlation $\left \langle \hat{a}_S \hat{a}_I \right \rangle = c/2$, but zero phase-insensitive cross correlation. 

As will be seen later, the phase-sensitive correlation provides the signature of target presence in quantum illumination. Unfortunately, this correlation is the more difficult of the two to detect as it does not give rise to fringes in second order interference. A receiver design based on interfering the idler field with radiation received from the target vicinity must include some additional element to produce fringes, such as a phase conjugation \cite{guha2009gaussian, jo2020quantum}. An alternative approach is to modify the two-mode squeezed vacuum state to obtain a non-zero phase-insensitive cross correlation which may then take over as the signature of target presence given an appropriate receiver design. To that end, the effect of phase-sensitive, parametric amplification of the idler field is now considered.

\section{The Parametrically-Amplified Idler}
\label{AmplifiedIdler}
Parametric or phase-sensitive amplification is a process that boosts one field quadrature while attenuating the other \cite{gardiner2004quantum}. In terms of input quadrature operators $\hat{q}_\text{in}$ and $\hat {p}_\text{in}$, parametric amplification (with appropriately chosen phase) affects a transformation giving output operators of the form
\begin{align}
\hat{q}_\text{out} = G \hat{q}_\text{in}, \\
\hat{p}_\text{out} = G^{-1}\hat{p} _\text{in}
\end{align}
where $G$ is the gain. It is assumed that the amplifier bandwidth is matched to the bandwidth $W$ of the idler beam. Applying this transformation to the single idler mode of Eq. (\ref{TMSVcov}) produces an output characterized by the covariance matrix
\begin{align}
\mbox{\boldmath$V$} = \frac{1}{2}
\begin{pmatrix} 
\nu  & 0 & Gc  & 0 \\
0 & \nu  & 0 & -G^{-1}c  \\
Gc   & 0  & G^2 \nu &0 \\
0 & -G^{-1}c  & 0& G^{-2} \nu  
\end{pmatrix}
\label{SIcov}
\end{align}
The important consequence of parametric amplification of the idler is that it breaks the equality in magnitude of the off-diagonal terms so that Eq. (\ref{PICC}) now gives a phase-insensitive cross correlation $\left \langle \hat{a}_S^\dagger \hat{a}_I \right \rangle = \left(G - G^{-1} \right) c/4$, non-zero  for any gain greater than unity. Moreover, increasing amplification boosts the correlation from zero without requiring an increase in the mean transmitted signal energy, $MN_S$. As shown below, the non-zero phase-insensitive cross correlation provides a measurable signature for target detection using second order interference.

It is important to note that other possible operations that might be performed on the idler modes, such as phase-insensitive amplification or projective measurement, can destroy the entanglement between the idler and the signal modes. The signal-idler state then becomes a classical state and the illumination is no longer `quantum', strictly speaking. In contrast, phase-sensitive amplification of the idler does not destroy the entanglement. The covariance matrix of Eq. \ref{SIcov} represents a non-separable state. This fact can be verified by the positive partial transpose test for non-separability. This test requires a symplectic eigenvalue less than 1/2 for the covariance matrix that results from a mirror reflection of the momentum operator of one mode \cite{adesso2007entanglement}. A straightforward calculation shows that for Eq. \ref{SIcov}, the relevant symplectic eigenvalue has the value $2^{-1}\left( \sqrt{N_S} + \sqrt{N_S+1} \right)^{-2}$ which is less than $1/2$ for any non-zero value of $N_S$, and  independent of the amount of amplification of the idler. Thus a two-mode squeezed vacuum with a parametrically amplified idler constitutes a true quantum state for illuminating a target as described in the section that follows.

\section{Quantum Illumination}
\label{QuantumIllumination}
Target detection using quantum illumination is a binary hypothesis test where a quantum state interrogates a vicinity of interest to decide whether a target is absent (hypothesis $H_0$) or present (hypothesis $H_1$). The quantum state contains idler modes which are retained and signal modes which are transmitted. The radiation received from the vicinity of interest consists of either background thermal modes (under $H_0$) or a combination of background and weakly reflected signal modes (under $H_1$). In the latter case, noise and loss destroy the entanglement between signal and idler, but a remnant of the correlation remains that can be a stronger signature of target presence than would be achievable using any classical state \cite{tan2008quantum}.

Under hypothesis $H_0$, each mode $\hat{a}_R$ of the received field is simply a thermal state $\hat{a}_B$ with mean photon number $N_B$. Then the two-mode covariance matrix for a received mode and an idler mode is
\begin{align}
\mbox{\boldmath$V$}^0_{\text{R,I}} = \frac{1}{2}
\begin{pmatrix} 
\omega  & 0 & 0  & 0 \\
0 & \omega  & 0 & 0  \\
0   & 0  & G^2 \nu &0 \\
0 & 0  & 0& G^{-2} \nu  
\end{pmatrix}
\label{QIcov_0}
\end{align}
where $\omega = 2N_B +1$. Under hypothesis $H_1$, the target weakly reflects the signal modes with reflectance $\kappa << 1$, so a received mode is a mix of signal $\hat{a}_S$ and background $\hat{a}_B$ such that $\hat{a}_R = \sqrt{\kappa} \hat{a}_S + \sqrt{1- \kappa}\hat{a}_B$, where the background mean photon number is set to $N_B/(1-\kappa)$ to avoid a passive target signature. In this case, the covariance matrix for a pair of received and idler modes is
\begin{align}
\mbox{\boldmath$V$}^1_{\text{R,I}} = \frac{1}{2}
\begin{pmatrix} 
\gamma  & 0 & \sqrt{\kappa}Gc  & 0 \\
0 & \gamma  & 0 & -\sqrt{\kappa}G^{-1}c  \\
\sqrt{\kappa}Gc   & 0  & G^2 \nu &0 \\
0 & -\sqrt{\kappa}G^{-1}c  & 0& G^{-2} \nu  
\end{pmatrix},
\label{QIcov_1}
\end{align}
where $\gamma = 2\kappa N_S + \omega$. The off diagonal elements of this matrix represent the remnant of the correlations of the transmitted quantum state. For realistic values of $N_S$, $N_B$, and $\kappa$, these correlations are too small to constitute quantum entanglement \cite{shapiro2020quantum}. But if they are measurable, they may serve as the signature of target presence. Importantly, the inequality of  magnitudes of the off-diagonal elements implies a non-zero phase-insensitive cross correlation that may be detected simply as second order interference.

A standard apparatus for measuring second order interference consists of a beam splitter with a photon-counting detector at each output port (see Fig. \ref{fig:Schematic Diagram}(a)). The difference in the photon counts is a measure of second order interference. The action of a 50-50 beam splitter combining two modes with covariance matrix $\mbox{\boldmath$V$}$ is represented by the symplectic matrix
\begin{align}
\mbox{\boldmath$B$} = \frac{1}{\sqrt{2}}
\begin{pmatrix} 
1  & 0 &1   & 0 \\
0 &1  & 0 & 1   \\
1    & 0  &-1  &0 \\
0 &1  & 0&-1   
\end{pmatrix}
\end{align}
where the covariance matrix of the modes at the output ports, represented by the operators $\hat{a}_+$ and $\hat{a}_-$,  is 
\begin{align}
\mbox{\boldmath$V$}_\pm = \mbox{\boldmath$B$}\mbox{\boldmath$V$}\mbox{\boldmath$B$}^T.
\end{align}
 Thus, combining the received field with the idler field on a beam splitter gives the covariance matrix
\begin{align}
\mbox{\boldmath$V$}^0_{\pm} = \frac{1}{2}
\begin{pmatrix} 
\frac{\omega + G^2\nu}{2}  & 0 &\frac{\omega - G^{2}\nu}{2}   & 0 \\
0 & \frac{\omega + G^{-2}\nu}{2}   & 0 & \frac{\omega - G^{-2}\nu}{2}   \\
\frac{\omega - G^{2}\nu}{2}    & 0  &\frac{\omega + G^2\nu}{2}  &0 \\
0 & \frac{\omega - G^{-2}\nu}{2}   & 0& \frac{\omega + G^{-2}\nu}{2}   
\end{pmatrix}
\label{QIcov_0}
\end{align}
under hypothesis $H_0$, and 
\begin{align}
\mbox{\boldmath$V$}^1_{\pm} = \frac{1}{2} \begin{pmatrix} 
\frac{\gamma + G^2\nu+2\sqrt{\kappa}Gc}{2}  & 0 &\frac{\gamma - G^{2}\nu}{2}   & 0 \\
0 & \frac{\gamma +\frac{\nu}{G^2} -2\sqrt{\kappa}\frac{c}{G}}{2}  & 0 & \frac{\gamma - G^{-2}\nu}{2}   \\
\frac{\gamma - G^{2}\nu}{2}    & 0  &\frac{\gamma + G^2\nu-2\sqrt{\kappa}Gc }{2} &0 \\
0 & \frac{\gamma - G^{-2}\nu}{2}   & 0& \frac{\gamma + \frac{\nu}{G^2} +2\sqrt{\kappa}\frac{c}{G}}{2} 
\end{pmatrix}
\label{QIcov_1}
\end{align}
under $H_1$. Photodetectors at the output ports measure $\hat{N}_+$ and $\hat{N}_-$, the photon number of the output modes $\hat{a}_+$ and $\hat{a}_-$, respectively. The difference between the counts is formed electronically.

The receiver counts photons from all modes on each output of the beam splitter. The photon count for any single mode is a non-Gaussian random variable, but all the counts are identically distributed. For large enough time-bandwidth product (i.e. number of modes) $M$, the central limit theorem implies that the total photon count difference  is Gaussian with mean and variance given by
\begin{align}
 \langle \hat{N}_i \rangle &= M\left( \langle \hat{ N}_{+,i} \rangle-\langle\hat{ N}_{-,i}\rangle  \right) 
\end{align}
and 
\begin{align}
\langle \Delta \hat{N}_i^2 \rangle = M \left(\left\langle \left(\hat{N}_{+,i} - \hat{N}_{-,i} \right)^2\right\rangle  -   \left\langle \hat{N}_{+,i} - \hat{N}_{-,i}  \right\rangle^2\right),
\end{align}
respectively, with $i =0,1$ referring to the corresponding hypothesis.

\begin{figure}[tbh]
\includegraphics{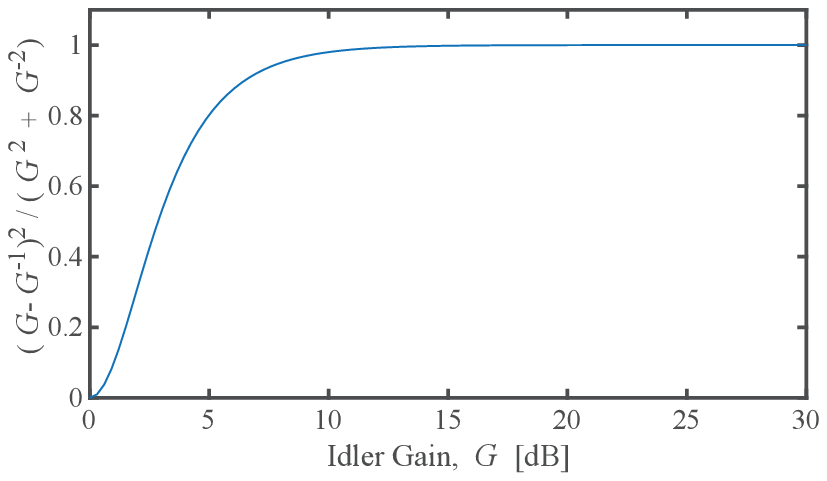}
\caption{(Color online) The gain dependent prefactor in the signal-to-noise ratio vanishes in absence of amplification but quickly rises to unity when amplification is applied.}
\label{fig:G_curve}
\end{figure}

From the Gaussian distributions for the two hypotheses, and setting the decision threshold to
\begin{align}
\frac{M\left(   \langle \hat{N}_0 \rangle\sqrt{\langle \Delta \hat{N}_1^2 \rangle}+ \langle \hat{N}_1 \rangle\sqrt{\langle \Delta \hat{N}_0^2 \rangle} \right) }{\sqrt{\langle \Delta \hat{N}_1^2 \rangle}+\sqrt{\langle \Delta \hat{N}_0^2 \rangle}},
\end{align}
the total error probability for equal probability discrimination is 
\begin{align}
P_{\text{SI}} = \frac{1}{2} \text{erfc} \left( \sqrt{M \cdot \text{SNR}_{QI}} \right)
\label{P_SI}
\end{align}
where
\begin{align}
\text{SNR}_\text{QI} = \frac{\left ( G - G^{-1} \right )^2}{\left (G^2 + G^{-2} \right )} \times\frac{ \kappa c^2 }{\left ( \sqrt{\gamma \nu + \kappa c^2} + \sqrt{\nu \omega} \right )^2}
\label{SNR_QISI}
\end{align}
is the single-mode-pair signal-to-noise ratio \cite{footnote1}. Notably, the gain-dependent leading factor is zero for unity gain. This is a direct reflection of the fact that the phase-insensitive cross correlation is zero without amplification. The leading factor is approximately unity for any gain $G>>1$ as seen in Fig. \ref{fig:G_curve}. Thus amplification of the idler beyond about 15 dB produces only a negligible improvement in the signal-to-noise ratio.

The significance of Eq. (\ref{SNR_QISI}) becomes apparent when compared to target detection schemes based on classical states. In the next section, the performance  of quantum illumination with a parametrically amplified idler is shown to exceed that of a  classical-state detection scheme that asymptotically matches the best possible classical-state scheme.

\section{Advantage Over Classical-State Target Detection}
\label{QuantumAdvantage}
In this section, quantum illumination with a parametrically amplified idler is tested against an important classical-state benchmark. The benchmark system is a specific detection scheme with an error exponent that approaches the optimal value for coherent states in the strong background ($N_B>>1$) regime. This system uses a multi-mode coherent state as the signal and coherently integrated, mode-by-mode homodyne detection as the receiver. For brevity, this detection scheme will be referred to as \em coherent-state homodyne detection\em. 

Assuming the signal pulse contains $M$ modes each with mean photon number $N_S$ (to match the energy in the quantum illumination signal), the total  probability of error for detection with equal prior probabilities is of the form of Eq. (\ref{P_SI}) but with the single-mode signal-to-noise ratio \cite{shapiro2009quantum, guha2009gaussian}
\begin{align}
 \text{SNR}_{\text{CSH}} = \frac{\gamma - \omega}{4\omega} = \frac{\kappa N_S}{4N_B+2}.
 \label{SNR_CSH}
\end{align}
When $N_B >> 1$, $ \text{SNR}_{\text{CSH}} \approx \kappa N_S/(4N_B)$ which approximates the quantum Chernoff bound error exponent for optimum reception using a coherent state \cite{tan2008quantum}. Importantly, this bound has been shown also to be optimal for all classical state signals. Thus, coherent-state homodyne detection is significant as a benchmark insofar as it is a structured receiver that approaches optimal performance in this regime \cite{shapiro2009quantum, guha2009gaussian, Karsa2020Quantum}. 

When comparing the single-mode signal-to-noise ratio in Eq. (\ref{SNR_CSH}) to Eq. (\ref{SNR_QISI}) assuming a strong background (i.e. $N_B>>1$), three regimes may be distinguished. First, consider the case where $N_S<<1$. This is the low-signal, strong-background regime originally identified as the domain where quantum illumination with a Gaussian state outperforms target detection using any classical state \cite{tan2008quantum}. Under these conditions, Eq. \ref{SNR_QISI} gives $\text{SNR}_\text{QI} \approx \kappa N_S/(2N_B)$, a 3 dB (i.e. a factor of 2) improvement over coherent-state homodyne detection. This result is comparable to the performance of phase conjugation and optical parametric amplifier receivers \cite{guha2009gaussian}.

 Next, let $1<<N_S<<N_B/\kappa$. Then Eq. \ref{SNR_QISI} gives $\text{SNR}_\text{QI} \approx \kappa N_S/(4N_B)$. Quantum illumination only matches optimal classical-state detection in this regime. Since typically it is assumed $\kappa<<1$, this regime can be quite broad. Given the relative difficulty of implementing quantum illumination, it is unlikely that it would have any practical advantage over a presumably much simpler classical-state scheme for applications limited to this regime. However, for applications requiring operation in both this regime and the previous one, it could be said that quantum illumination would provide performance equal to or better than any classical state scheme.

Finally, at even greater signal brightness levels where $N_S >> N_B/\kappa$, the signal-to-noise ratio saturates with $\text{SNR}_\text{QI} \approx 1/2$. In this regime, quantum illumination with or without an amplified idler is no longer competitive with classical-state target detection. It is difficult to imagine any rationale for using quantum illumination in this regime.

\begin{figure}[tbh]
\includegraphics{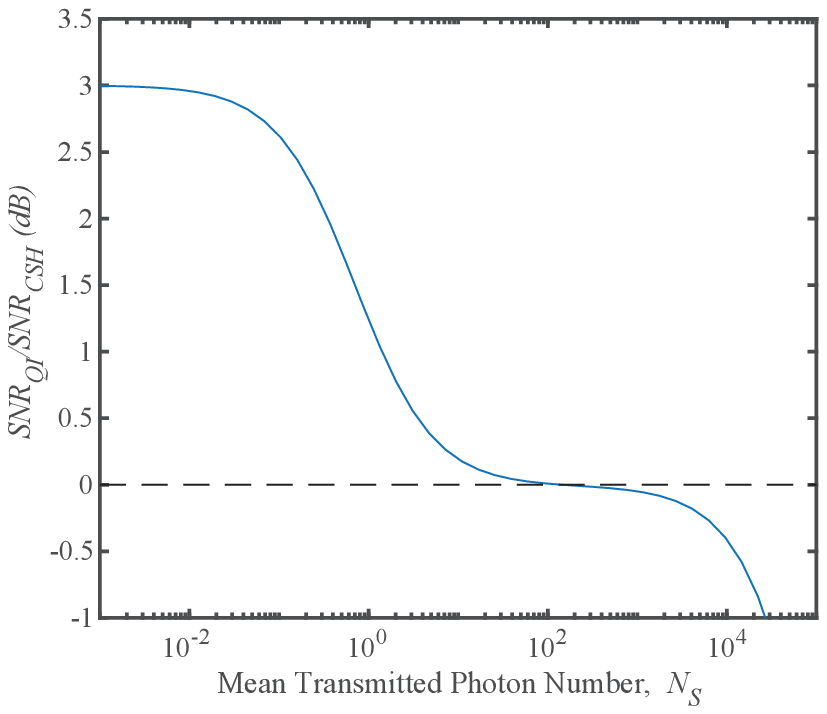}
\caption{(Color online) Ratio of single-mode signal-to-noise ratios for quantum illumination with a parametrically amplified idler and coherent-state homodyne detection with $G = 15$ dB, $N_B = 10^{2}$, and $\kappa = 10^{-3}$.}
\label{fig:QA_over_CSH}
\end{figure}

To illustrate these cases, the ratio of the signal-to-noise ratio for quantum illumination with a parametrically amplified idler to that of coherent-state homodyne detection  is plotted in Fig. \ref{fig:QA_over_CSH}  with a gain of $G= 15$ dB,  background photon number $N_B = 10^2$, and reflectance $\kappa= 10^{-3}$. All three regimes are apparent as well as the transitions from a 3 dB advantage to 0 dB around $N_S \approx 1$, and from 0 dB to a quantum disadvantage around $N_S \approx N_B/\kappa = 10^5$.

\section{Discussion}
\label{Discussion}
The introduction of parametric amplification to the idler beam has been shown to enable reception using a simple and familiar configuration of a beam splitter and photodetectors. Phase-sensitive amplification modifies the correlation structure of the original two-mode squeezed vacuum to produce a non-zero phase-insensitive cross correlation detectable in second order interference. The simple receiver of Fig. \ref{fig:Schematic Diagram}(b) enabled by parametrically amplifying the idler may be viewed as coming at a cost of increasing the complexity of the transmitter. This architecture trades a nonlinear process at the receiver, for example phase conjugation or sum frequency generation \cite{guha2009gaussian, zhuang2017optimum}, for a nonlinear process in the transmitter, phase-sensitive amplification. However, there may be cause to prefer the latter since it is reasonable to expect that the parameters of the idler entering the parametric amplifier can be more carefully controlled than those of received light entering a phase conjugator or other nonlinear element. 



\bibliography{QIwithSqueezedIdler}
\end{document}